\documentclass[10 pt, conference]{IEEEtran}

\usepackage{amsmath, amssymb}
\usepackage{graphicx}
\usepackage[sans]{dsfont}
\usepackage{stmaryrd}
\usepackage{mathrsfs}
\usepackage{subfigure}
\usepackage{color}
\usepackage{cite}
\usepackage{asymptote}
\usepackage{algorithm}
\usepackage{algorithmic}

\usepackage{flushend}

\usepackage{calligra}
\DeclareFontShape{T1}{calligra}{m}{n}{<->s*[2.2]callig15}{}
\DeclareMathAlphabet{\mathcalligra}{T1}{calligra}{m}{n}
\DeclareMathAlphabet{\mpzc}{OT1}{pzc}{m}{it}

\def\vect#1{\underline{#1}}

\newtheorem{thm}{Theorem}
\newtheorem{lem}[thm]{Lemma}

\newtheorem{prop}[thm]{Proposition}

\newtheorem{defn}[thm]{Definition}
\newtheorem{eg}[thm]{Example}

\newenvironment{dis}[1][Discussion:]{\begin{trivlist}
\item[\hskip \labelsep {\itshape #1} \hspace*{1mm}]}{\end{trivlist}}

\title{\LARGE {\bf Enhancing Binary Images of Non-Binary LDPC Codes}}
\author{
\authorblockN{Aman Bhatia, Aravind R. Iyengar, and Paul H. Siegel}
\authorblockA{University of California, San Diego, La Jolla, CA $92093-0401$, USA \\Email: \{a1bhatia, aravind, psiegel\}@ucsd.edu}
}

\newcommand{\twobibs}[2]{#2}

\begin{document}
\maketitle

\begin{abstract}
We investigate the reasons behind the superior performance of belief propagation decoding of non-binary LDPC codes over their binary images when the transmission occurs over 
the binary erasure channel. We show that although decoding over the binary image has lower complexity, it has worse performance owing to its 
larger number of stopping sets relative to the original non-binary code. We propose a method to find redundant parity-checks of the binary 
image that eliminate these additional stopping sets, so that we achieve performance comparable to that of the original non-binary LDPC code 
with lower decoding complexity.
\end{abstract}

\section{Introduction}
Low-density parity-check (LDPC) codes were introduced by Gallager \cite{gal_63_the_ldpc} and rediscovered in the $1990$'s. 
Davey and MacKay showed that non-binary LDPC (NB-LDPC) codes perform better than binary codes for the same bit length \cite{dav_98_itw_nbldpc}.
Despite their better performance, higher decoding complexity has limited the implementation of non-binary LDPC codes in real-world applications. Reduced complexity decoding is possible using an equivalent binary Tanner graph, called the binary image of the code.
There are many ways to construct binary images of non-binary codes - one way, referred to as basic binary image, is to represent 
each non-binary codeword by a binary vector of the same bit length. A recent result \cite{sav_09_itw_nbldpc} defines an extended 
binary image of longer bit length and establishes the equivalence of belief propagation (BP) decoding over it to BP decoding over the 
non-binary codes for the binary erasure channel (BEC). Based on this image, we compare NB-LDPC codes to their basic binary images
and suggest an algorithm to find redundant parity-checks to improve the performance of the basic binary images.

This paper is organized as follows. In Section \ref{sec_nbldpc}, we briefly describe the BP decoding of NB-LDPC codes. 
We define the basic and extended binary images in Section \ref{sec_binimg}. The superiority of BP 
over NB-LDPC code for BEC compared to its basic binary image is shown in Section \ref{sec_supnb} and 
the algorithm to bridge this gap in performance through addition of 
redundant parity-checks to the basic binary image is given in Section \ref{sec_enh}. We summarize our findings in Section \ref{sec_conc}.

\section{$\mathrm{GF}(2^{b})$-LDPC Codes} \label{sec_nbldpc}

An LDPC code $\mathcal{C}_{q}$ over $\mathrm{GF}(q)$ where $q=2^{b}$
is specified by a \emph{parity-check matrix} $\mathbf{H}_{q}$ of
dimension $m_{q}\times n_{q}$ whose elements are from $\mathrm{GF}(q)$,
where the number of non-zero elements in $\mathbf{H}_{q}$ is proportional
to $n_{q}$. The code is then defined as the \emph{nullspace} of $\mathbf{H}_{q}$,
i.e., \[
\mathcal{C}_{q}=\{\underline{X}\in\mathrm{GF}(q)^{n_{q}}:\mathbf{H}_{q}\otimes_{q}\underline{X}^{\mathsf{T}}=\underline{0}^{\mathsf{T}}\}\]
 where $\underline{X}$ is assumed to be a row-vector, $\underline{0}$
the all-zero row-vector, $\otimes_{q}$ denotes matrix multiplication
when matrix elements belong to $\mathrm{GF}(q)$, and $^{\mathsf{T}}$
denotes transposition. When the parity-check matrix $\mathbf{H}_{q}$
is full-rank, the code $\mathcal{C}_{q}$ is a vector-subspace of
$\mathrm{GF}(q)^{n_{q}}$ of \emph{dimension} $k_{q}=n_{q}-m_{q}$,
has a \emph{blocklength} $n_{q}$ symbols, and is therefore of \emph{rate}
$R=\frac{k_{q}}{n_{q}}=1-\frac{m_{q}}{n_{q}}$.

We associate with the matrix $\mathbf{H}_{q}$ a bipartite graph,
called the \emph{Tanner graph} $\mathcal{G}_{q}$, as follows. Corresponding
to the $j^{\text{th}}$ column of $\mathbf{H}_{q}$ is a \emph{variable}
node $V^q_{j}$, and corresponding to the $i^{\text{th}}$ row is a
\emph{check} node $C^q_{i}$, $i \in [m_q] \triangleq \{1, 2, \ldots, m_q\}, j \in [n_q]$.
Every non-zero entry $h_{i,j}$ of $\mathbf{H}_{q}$ corresponds
to an edge between $C^q_{i}$ and $V^q_{j}$ with label $h_{i,j}$.
We denote by $\mathcal{N}^v(j)$ the set of check nodes connected to
a given variable node $j \in [n_q]$, i.e., the neighbors of the $j^\text{th}$ variable node, and
by $\mathcal{N}^c(i)$ the set of variable nodes connected to a given
check node $i \in [m_q]$.
When the degree of every variable node and every check node in $\mathcal{G}_q$ 
is $d_{l}$ and $d_{r}$ respectively,  i.e.,
$|\mathcal{N}^v(j)| = d_l,{\ } |\mathcal{N}^c(i)| = d_r{\ }\forall{\ }i \in
[m_q], j \in [n_q]$, 
the LDPC code $\mathcal{C}_{q}$ is said to
be $(d_{l},d_{r})$-regular. It is easy to see that for a $(d_{l},d_{r})$-regular
code, the rate satisfies $R=1-\frac{m_{q}}{n_{q}}=1-\frac{d_{l}}{d_{r}}$.

\subsection*{Belief Propagation Decoding}
The messages passed around on the Tanner 
graphs in the BP decoder represent the {\it a posteriori}
probabilities of the symbols of $\mathrm{GF}(q)$. We will briefly
describe the BP decoding of non-binary LDPC codes over the binary
erasure channel (BEC) using the analogue of the \emph{peeling decoder},
which for the BEC is the same as the BP decoder over $\mathcal{G}_q$, which we denote as $\mathrm{BP}(\mathbf{H}_q)$.

We assume a fixed isomorphism $\Phi_B : \mathrm{GF}(q)\mapsto\mathrm{GF}(2)^{b}$
that preserves the addition operation defined on $\mathrm{GF}(q)$.
This isomorphism is used to map codewords $\underline{X} \in\mathcal{C}_{q}\subset\mathrm{GF}(q)^{n_{q}}$
to binary vectors $\underline{x} \in\mathrm{GF}(2)^{bn_{q}}$.
The vector $\underline{x}$ is transmitted over a BEC with erasure 
probability $\epsilon$ and let $\mathcal{E}$ be the index set of the 
erasures in the received word. At the receiver, the \textit{a priori}
set of eligible symbols $\mathcal{Q}_{j}^{(0)}$ for the $j^{\text{th}}$
variable node consists of symbols which fit the received binary sequence
corresponding to the $j^{\text{th}}$ transmitted symbol. Thus $\mathcal{Q}_{j}^{(0)}$
has $2^{a}$ symbols if $a$ out of the $b$ bits in the received
binary sequence corresponding to the $j^{\text{th}}$ transmitted
symbol are erased. The peeling decoder updates these sets iteratively
by exchanging messages between variable and check nodes. For the BEC, 
the messages passed around iteratively can be represented by sets of 
eligible symbols. Let $\mathcal{Q}_{i,j}^{(\ell)}$ denote the set of eligible 
symbols sent by the $i^{\text{th}}$ check node to the $j^{\text{th}}$ 
variable node in the $\ell^\text{th}$ iteration. Then, the peeling decoder performs the following steps iteratively for $\ell \geq 1$
\begin{itemize}
\item {\it Check node processing}: For check node $C^q_{i}, i \in [m_q]$,
\[
\mathcal{Q}_{i,j}^{(\ell)} = \sum_{j'\in\mathcal{N}^c(i)\backslash\{j\}}h_{i,j'}\mathcal{Q}_{j'}^{(\ell - 1)} \quad\forall\, j\in\mathcal{N}^c(i). 
\]
\item {\it Variable node processing}: For variable node $V^q_{j}, j \in [n_q]$,
\[
\mathcal{Q}_{j}^{(\ell)} = \mathcal{Q}_{j}^{(\ell - 1)} \cap \Big(\bigcap_{i\in\mathcal{N}^v(j)}h_{i,j}^{-1}\mathcal{Q}_{i,j}^{(\ell)} \Big).
\]
\end{itemize}
where, if $h \in \mathrm{GF}(q)$ and $\mathcal{Q}, \mathcal{Q}_{1}, \mathcal{Q}_{2} \subseteq \mathrm{GF}(q)$, we define 
\begin{align*}
h \mathcal{Q} &\triangleq \{ h a \, \vert \, a \in \mathcal{Q}\} \\
\mathcal{Q}_{1} + \mathcal{Q}_{2} &\triangleq \{a_1 + a_2 \, \vert \, a_1 \in \mathcal{Q}_{1}, a_2 \in \mathcal{Q}_{2} \}.
\end{align*}

The decoder stops when $|\mathcal{Q}_{j}^{(\ell)}| = |\mathcal{Q}_j^{(\ell - 1)}|{\ }\forall{\ }j \in [n_q]$. 
Decoding is successful if $|\mathcal{Q}_{j}^{(\ell)}|=1{\ }\forall{\ } j \in [n_q]$ for some $\ell \in \mathbb{N}$.
It should be noted that any set of eligible symbols ($\mathcal{Q}_{j}^{(\ell)}$ or $\mathcal{Q}_{i,j}^{(\ell)}$)
is a coset of a vector subspace of $\mathrm{GF}(q)$ \cite{rat_05_comproc_nbldpc}.

\section{Binary Images} \label{sec_binimg}
Since codes defined over $\mathrm{GF}(2^b)$ can be written as a collection of 
vectors over $\mathrm{GF}(2)$, we can consider the \emph{binary images} of these 
codes. Further, since BP decoding for $\mathrm{GF}(q)$ has a high
computational complexity, we can make use of the binary images to decode the non-binary
code.

\subsection{Basic Binary Image}
The isomorphism $\Phi_B$ defined in Section \ref{sec_nbldpc} maps any row-vector
of symbols $\underline{X}=(X_1, \ldots, X_{n_q}) \in \mathrm{GF}(q)^{n_q}$
to a binary row-vector $\Phi_B(\underline{X})\triangleq(\Phi_B(X_1), \ldots, \Phi_B(X_{n_q}))$
of length $bn_q$. The \textit{basic binary image} of $\mathcal{C}_{q}$
is defined as the set $\mathcal{C}_B \subset \mathrm{GF}(2)^{bn_q}$ where 
$\mathcal{C}_B = \{ \Phi_B(\vect{X}) : \vect{X} \in \mathcal{C}_q \}$.
As described earlier, the codewords of $\mathcal{C}_B$ are the ones that are transmitted over the BEC. 
Note that there are $\prod_{l=0}^{b-1}\left(2 ^ b - 2 ^ l\right)$ possible choices for the isomorphism $\Phi_B$.

Once fixed, the isomorphism $\Phi_B$ identifies a mapping $\Psi_m : \mathrm{GF}(q) \mapsto \mathcal{M}_{b}$,
where $\mathcal{M}_{b}$ is the collection of all invertible matrices
over $\mathrm{GF}(2)$ of size $b \times b$, defined below. For any $h \in \mathrm{GF}(q)$,
\[
\Psi_m(h) = 
\left(\begin{array}{c}
\Phi_B\left(h\times\Phi_B^{-1}\left(\underline{u}_{1,b}\right)\right)\\
\Phi_B\left(h\times\Phi_B^{-1}\left(\underline{u}_{2,b}\right)\right)\\
\vdots\\
\Phi_B\left(h\times\Phi_B^{-1}\left(\underline{u}_{b,b}\right)\right)
\end{array}
\right)^{\mathsf{T}}
\]
where $\underline{u}_{i,b}$ denotes the binary row-vector of length $b$ with the element $i$ as $1$ 
and other elements equal to $0$, i.e., the \emph{unit vector} along the $i^\text{th}$ dimension. 
The set $\left\{ \Psi_m(h):h\in\mathrm{GF}(q)\right\} \subset\mathcal{M}_{b}$
constitutes a field under matrix addition and matrix multiplication
operations over $\mathrm{GF}(2)$, i.e., for $h_{1},h_{2}\in\mathrm{GF}(q)$, $\Psi_m(h_1 + h_2) = \Psi_m(h_1) \oplus_2 \Psi_m(h_2)$ 
and $\Psi_m(h_1h_2) = \Psi_m(h_1) \otimes_2 \Psi_m(h_2)$. Also, it is easy to see that for any $h, x \in \mathrm{GF}(q)$, 
$\Psi_m(h) \otimes_2 \Phi_B(x) ^ {\mathsf{T}} = \Phi_B(hx) ^ {\mathsf{T}}$.

The mappings $\Phi_B$ 
and $\Psi_m$ allow us to identify $\mathcal{C}_B$ as a 
linear block code with parity-check matrix $\mathbf{H}_B$ that is obtained by replacing each element
of $\mathbf{H}_q$ by its image given by $\Psi_m$. We could perform BP on the Tanner graph of $\mathbf{H}_B$, 
denoted by $\mathrm{BP}(\mathbf{H}_B)$, instead of the more complex $\mathrm{BP}(\mathbf{H}_q)$.
However this has worse performance in general due to the large number of short cycles introduced by 
$\Psi_m(h)$'s in the parity-check matrix $\mathbf{H}_B$. In particular, for the BEC, we shall show later that 
this bad performance can be attributed to a larger number of stopping sets of $\mathbf{H}_B$.

\subsection{Extended Binary Image}
The observation made at the end of the previous subsection leads us to the question  
whether it is possible to design a binary image of a non-binary code that matches the 
performance of the non-binary code. This question was answered in the affirmative by 
Savin in \cite{sav_09_itw_nbldpc} for the BEC. We briefly describe the construction of 
this code, called the \emph{extended binary image}.

We define a mapping $\Phi_E : \mathrm{GF}(q) \mapsto \mathrm{GF}(2)^{2^b - 1}$
such that each symbol of $\mathrm{GF}(q)$ is mapped to a codeword
of the \emph{simplex code} $\mathcal{C}_S(b)$ of length $2^b - 1$, i.e., the dual of the 
Hamming code of blocklength $2^b - 1$. For $X \in \mathrm{GF}(q)$,
\[
\Phi_E(X) \triangleq \Phi_B(X) \otimes_2 \mathbf{G}_S(b)
\]
where $\mathbf{G}_S(b)$ is the generator matrix of the simplex code of size $b \times (2^b - 1)$. 
By linearity of $\Phi_B$, $\Phi_E(X_1 + X_2) = \Phi_E(X_1) \oplus_2 \Phi_E(X_2)$.
Note that the columns of $\mathbf{G}_S(b)$ are all vectors of weight $1 \leq w \leq b$. 
We will let the $i^\text{th}$ column of $\mathbf{G}_S(b)$ be $(g_{1, i}, g_{2, i}, \ldots, g_{b, i})^{\mathsf{T}}$ 
where $i = \sum_{j = 1}^bg_{j, i}2^{j - 1}$ for $i \in [2^b - 1]$.
We use the mapping $\Phi_E$ to map row-vector of symbols $\vect{X}=(X_1, \ldots, X_{n_q}) \in \mathrm{GF}(q)^{n_q}$
to a binary row-vector $\Phi_E(\vect{X})\triangleq(\Phi_E(X_1), \ldots, \Phi_E(X_{n_q}))$
of length $(2^b - 1)n_q$. The ordering of the columns of $\mathbf{G}_S(b)$ implies that for $i \in [n_q]$, $j \in [b]$
\begin{equation}
(\Phi_B(X_i))_j = (\Phi_E(X_i))_{2^{j-1}} \label{eq-not-trans},
\end{equation} where we write $(\Phi_B(X))_e$ to denote the $e^\text{th}$ 
element of the vector representing $\Phi_B(X)$ for $X \in \mathrm{GF}(q)$.
The \emph{extended binary image} of $\mathcal{C}_q$ is defined as the set $\mathcal{C}_E \subset \mathrm{GF}(2)^{(2^b-1)n_q}$
where 
$\mathcal{C}_E = \{ \Phi_E(\vect{X}) : \vect{X} \in \mathcal{C}_q \}$.
For a fixed $\Phi_E$, one can 
define a mapping $\Psi_M : \mathrm{GF}(q) \mapsto \mathcal{M}_{2^b - 1}$
such that
\[
\Psi_M(h) \otimes_2 \Phi_E(X)^{\mathsf{T}} = \Phi_E(hX)^{\mathsf{T}}{\ }\forall{\ }h, X \in \mathrm{GF}(q).
\]
Then the following can be shown.
\begin{lem}
\label{lem:phi-M-is-permutation}
$\Psi_M(h)$ is a permutation matrix for all $h \in \mathrm{GF}(q)$.
\end{lem}
The matrix $\mathbf{H}_{PM}$ of dimension $m_q(2^b - 1) \times n_q(2^b - 1)$
is defined as one obtained by replacing each element of $\mathbf{H}_{q}$
by its image under $\Psi_M$. Then, the Tanner graph of 
$\mathbf{H}_{PM}$ is a \emph{graph cover} \cite{ric_08_bok_mct} of $\mathcal{G}_q$  with all edge labels equal to $1$.
It is easy to show that the extended binary image is 
\[
\mathcal{C}_E = \{\underline{x} \in \mathrm{GF}(2)^{(2^b - 1)n_q} : \mathbf{H}_{PM} \otimes_2 \underline{x}^{\mathsf{T}} = \underline{0}^{\mathsf{T}}\} \cap \mathcal{C}_S(b)^{n_q}
\]
where
\[
\mathcal{C}_S(b)^{n_q} = \underbrace{\mathcal{C}_S(b) \times \ldots \times \mathcal{C}_S(b)}_{n_q} \subset \mathrm{GF}(2)^{(2^b - 1)n_q}.
\]
The overall parity-check matrix for the extended binary code is therefore of the form
\[
\mathbf{H}_E = \left(
\begin{array}{c}
\mathbf{H}_{PM} \\
\mathbf{I}_{n_q} \odot \mathbf{H}_S(b)
\end{array}
\right)
\]
where $\mathbf{I}_{n_q}$ is the $(n_q \times n_q)$ identity matrix, $\mathbf{H}_S(b)$ is the 
parity-check matrix of the simplex code and $\odot$ represents the \emph{Kronecker product}. 

As a consequence of Equation \eqref{eq-not-trans}, transmitting the codewords of $\mathcal{C}_B$ amounts to 
transmitting bits indexed by the set 
$\mathcal{I}_t = \{ i \in [(2^b-1)n_q] : i = 2^j \mod (2^b-1) , j \in \{0, 1, \ldots, b-1 \} \}$ of 
the corresponding codewords in $\mathcal{C}_E$ and puncturing the rest. 
We will let $\mathcal{I}_p = [(2^b-1)n_q] \setminus \mathcal{I}_t$ denote the index set of punctured bits in $\mathcal{C}_E$.
For ease of notation, we will define $x_{i,j}^B \triangleq (\Phi_B(X_i))_j$ and $x_{i,j}^E \triangleq (\Phi_E(X_i))_{j}$.
\begin{eg} \label{eg_1}
Let us consider a code over $\mathrm{GF}(8)$ given by the following
parity-check matrix $\mathbf{H}_q = (\alpha \quad \alpha^2 \quad 1)$, where $\alpha$ is a \emph{primitive element} of $\mathrm{GF}(8)$ with the \emph{primitive polynomial} being $x^3 + x + 1 = 0$. The code may be represented as solutions to the following linear constraint $\alpha X_1 + \alpha^2X_2 + X_3 = 0$.
\begin{table}[!h]
\centering
\caption{\label{tab:eg-phib}$\Phi_B:\mathrm{GF}(8)\mapsto\mathrm{GF}(2)^{3}$}
\begin{tabular}{cc}
$X\in\mathrm{GF}(8)$ & $\Phi_B(X)$\tabularnewline
\hline
\hline 
$0$ & $\left(0,0,0\right)$\tabularnewline
$\alpha$ & $\left(0,0,1\right)$\tabularnewline
$\alpha^{3}$ & $\left(0,1,0\right)$\tabularnewline
$\alpha^{6}$ & $\left(1,0,0\right)$\tabularnewline
$\alpha^{2}$ & $\left(1,1,1\right)$\tabularnewline
$\alpha^{4}$ & $\left(1,1,0\right)$\tabularnewline
$\alpha^{5}$ & $\left(1,0,1\right)$\tabularnewline
$\alpha^{7}$ & $\left(0,1,1\right)$\tabularnewline
\hline
\end{tabular}
\end{table}
Table~\ref{tab:eg-phib} gives the chosen mapping $\Phi_B$. 
Then the parity equations for the basic binary code can be written as 
\begin{align}
&\left(\begin{array}{ccc}
0 & 1 & 1\\
1 & 1 & 1\\
1 & 0 & 1
\end{array}\right)
\left(\begin{array}{c}
x_{1, 1}^B\\
x_{1, 2}^B\\
x_{1, 3}^B\end{array}\right)
+
\left(\begin{array}{ccc}
0 & 1 & 0\\
0 & 0 & 1\\
1 & 1 & 0
\end{array}\right)
\left(\begin{array}{c}
x_{2, 1}^B\\
x_{2, 2}^B\\
x_{2, 3}^B
\end{array}\right)
\notag \\
&\qquad\qquad+
\left(\begin{array}{ccc}
1 & 0 & 0\\
0 & 1 & 0\\
0 & 0 & 1
\end{array}\right)
\left(\begin{array}{c}
x_{3, 1}^B\\
x_{3, 2}^B\\
x_{3, 3}^B
\end{array}\right)
=
\left(\begin{array}{c}
0\\
0\\
0
\end{array}\right),\text{ i.e.,} \notag
\end{align}
\begin{align}
(x_{1, 2}^B + x_{1, 3}^B) + x_{2, 2}^B + x_{3, 1}^B &= 0, \label{eq:eg-bb-parity-equation-1}\\
(x_{1, 1}^B + x_{1, 2}^B + x_{1, 3}^B) + x_{2, 3}^B + x_{3, 2}^B &= 0, \label{eq:eg-bb-parity-equation-2}\\
(x_{1, 1}^B + x_{1, 3}^B) + (x_{2, 1}^B + x_{2, 2}^B) + x_{3, 3}^B &= 0. \label{eq:eg-bb-parity-equation-3}
\end{align}
The parity-check equation for the extended binary image can be written as
\[
\Psi_M(\alpha)\Phi_E(X_1) + \Psi_M(\alpha^{2})\Phi_E(X_2) + \Psi_M(1)\Phi_E(X_3) = 0.
\]
Thus, the seven binary parity-check equations represented by the equation 
above are represented by $\mathbf{H}_{PM}$ given by \vspace{2mm}
\begin{center}
\begin{tabular}{c|@{}c@{} c@{} c@{} c@{} c@{} c@{} c@{}|@{}c@{} c@{} c@{} c@{} c@{} c@{} c@{}|@{}c@{} c@{} c@{} c@{} c@{} c@{} c@{}|}
 & \multicolumn{7}{c|}{$(\Phi_E(X_1))_j$} & \multicolumn{7}{c|}{$(\Phi_E(X_2))_j$} & \multicolumn{7}{c|}{$(\Phi_E(X_3))_j$}\tabularnewline
$j$ & 1\; & 2\; & 3\; & 4\; & 5\; & 6\; & 7\; & 1\; & 2\; & 3\; & 4\; & 5\; & 6\; & 7\; & 1\; & 2\; & 3\; & 4\; & 5\; & 6\; & 7\; \tabularnewline
\hline
$P_{1}$ &  &  &  &  &  & 1 &  &  & 1 &  &  &  &  &  & 1 &  &  &  &  &  & \tabularnewline
$P_{2}$ &  &  &  &  &  &  & 1 &  &  &  & 1 &  &  &  &  & 1 &  &  &  &  & \tabularnewline
$P_{3}$ & 1 &  &  &  &  &  &  &  &  &  &  &  & 1 &  &  &  & 1 &  &  &  & \tabularnewline
$P_{4}$ &  &  &  &  & 1 &  &  &  &  & 1 & &  &  &  &  &  &  & 1 &  &  & \tabularnewline
$P_{5}$ &  &  & 1 & &  &  &  & 1 &  &  &  &  &  &  &  &  &  &  & 1 &  & \tabularnewline
$P_{6}$ &  & 1 &  &  &  &  &  &  &  &  &  &  &  & 1 &  &  &  &  &  & 1 & \tabularnewline
$P_{7}$ &  &  &  & 1 &  &  &  &  &  &  &  & 1 &  &  &  &  &  &  &  &  & 1\tabularnewline
\end{tabular}
\end{center} \vspace{2mm}
Here, equation $P_{1}$ is $x_{1, 6}^E + x_{2, 2}^E + x_{3, 1}^E = 0$
which together with the simplex constraints $x_{1, 6}^E = x_{1, 2}^B + x_{1, 3}^B$, $x_{2, 2}^E = x_{2, 2}^B$ 
and $x_{3, 1}^E = x_{3, 1}^B$  can be written as Equation
\eqref{eq:eg-bb-parity-equation-1}. Similarly, equations $P_{2}$
and $P_{4}$ correspond to Equations \eqref{eq:eg-bb-parity-equation-2}
and \eqref{eq:eg-bb-parity-equation-3} respectively. Equations $P_{3},P_5, P_6,P_{7}$
are all linear combinations of the above equations. \end{eg}

In \cite{sav_09_itw_nbldpc}, it was shown that the extended binary image was equivalent to the non-binary 
code from which it was constructed. But this equivalence was established for the BEC with BP updates at the 
parity-check nodes and ML updates at the simplex nodes. In order to achieve this equivalence with BP, we will 
assume that the parity-check matrices representing the simplex codes have redundant parity-checks assuring 
ML performance with BP. For the simplex codes, this is achieved by forming the parity-check matrix with rows 
of weight $3$ \cite{han_08_isit_mlred}. Hence, $\mathbf{H}_S(b)$ is the parity-check matrix of size 
$(2^b - 1)(2^{b - 1} - 1) \times (2^b - 1)$ for the simplex code. We will denote BP over the Tanner graph of 
$\mathbf{H}_E$ with $\mathbf{H}_S(b)$ having redundant checks and when only bits indexed by $\mathcal{I}_t$ 
are transmitted as $\mathrm{BP}(\mathbf{H}_E)$.

\section{Superiority of $\mathrm{BP}(\mathbf{H}_q)$ over $\mathrm{BP}(\mathbf{H}_B)$} \label{sec_supnb}
From the construction of the basic binary image and the extended binary image, it is clear that for each 
parity-check equation in $\mathbf{H}_q$, there are $b$ parity-check equations in $\mathbf{H}_B$ and 
$(2^b - 1)$ parity-check equations in the $\mathbf{H}_{PM}$ portion of $\mathbf{H}_E$.
\begin{lem} \label{lem:EB-has-BB-eqns}
For each row $\vect{h}_q$ in the parity-check matrix $\mathbf{H}_q$, 
the parity-check matrix $\mathbf{H}_E$ contains 
all non-trivial linear combinations of the $b$ rows corresponding to $\vect{h}_q$ in the parity-check matrix $\mathbf{H}_B$.
\end{lem}
\begin{dis}
First, note that the extra variables in the extended binary 
image were defined such that all non-trivial linear combinations of the transmitted bits were represented by a variable 
and these linear relations were maintained through the use of the simplex code for each non-binary symbol. Second, 
observe that by using the permutation matrices, we obtained a set of $(2^b - 1)$ parity-check equations each involving 
exactly one of the variables from each simplex codeword. Putting these two together, we can show the above result.
\end{dis}

Since it is known that the performance of $\mathrm{BP}(\mathbf{H}_q)$ is the same as that of 
$\mathrm{BP}(\mathbf{H}_E)$, it suffices to show that $\mathrm{BP}(\mathbf{H}_E)$  has superior performance in 
comparison with $\mathrm{BP}(\mathbf{H}_B)$. The performance of codes over the BEC with BP decoding is determined 
by the \emph{stopping sets} of the Tanner graph of their parity-check matrix \cite{di_02_tit_stopset}. 
By an \emph{erasure pattern} $\mathcal{E}$, we mean the index set of the erasures in a vector.
For a binary parity-check matrix $\mathbf{H}$, let $\beta(\mathbf{H},\mathcal{E})$ 
denote the erasure pattern obtained at the end of decoding the received word with erasure pattern 
$\mathcal{E}$ using BP on the Tanner graph of $\mathbf{H}$. Then a stopping set is an erasure 
pattern $\mathcal{E}$ such that $\beta(\mathbf{H}, \mathcal{E}) = \mathcal{E}$. Let 
$\mathbb{S}(\mathbf{H}) = \{\mathcal{E} : \beta(\mathbf{H}, \mathcal{E}) = \mathcal{E}\}$ be 
the set of stopping sets of $\mathbf{H}$. Notice that this condition is the same as the graph-theoretic 
requirement that every check node neighbor of variables in the Tanner graph of $\mathbf{H}$ indexed by 
$\mathcal{E}$ be connected to these variables at least twice. However, with the extended binary image, we 
are interested in erasure patterns only among the transmitted variables. Hence we do 
not have a corresponding graph-theoretic definition of stopping sets of $\mathrm{BP}(\mathbf{H}_E)$. 

\begin{defn}[Stopping Sets of $\mathrm{BP}(\mathbf{H}_E)$]
Consider BP decoding of the extended binary code over the BEC with the parity-check matrix $\mathbf{H}_E$. 
A stopping set, $\mathcal{E}$, is a subset of the index set of 
transmitted bits such that BP decoding can recover no transmitted bit in $\mathcal{E}$, i.e.,
\[
\mathcal{E} \subset \mathcal{I}_t \text{ such that } \beta(\mathbf{H}_E,\mathcal{E} \cup \mathcal{I}_p) \cap \mathcal{I}_t = \mathcal{E}.
\]
\end{defn}
\begin{dis}
Let us denote by $\mathbb{S}_E$ the set of all stopping sets of the extended binary image as defined above.
Let $\mathbb{S}(\mathbf{H}_E)$ denote the set of all stopping sets of the code with parity-check 
matrix $\mathbf{H}_E$ assuming all bits are transmitted. Define
\[
\mathbb{S}^t(\mathbf{H}_E) = \{\mathcal{S} \cap \mathcal{I}_t : \mathcal{S} \in \mathbb{S}(\mathbf{H}_E)\}.
\]
Then, $\mathbb{S}_E = \mathbb{S}^t(\mathbf{H}_E)$. This is because  
for each $\mathcal{S} \in \mathbb{S}^t(\mathbf{H}_E)$, 
$\beta(\mathbf{H}_E,\mathcal{S})$ is
the maximal stopping set $\mathcal{S}^\prime \in \mathbb{S}(\mathbf{H}_E)$ such that 
$\mathcal{S}^\prime \cap \mathcal{I}_t = \mathcal{S}$. Note that for the basic binary image, the set of all stopping sets is 
just the normal definition $\mathbb{S}_B = \mathbb{S}(\mathbf{H}_B)$ since every bit of the basic binary image is 
transmitted.
\end{dis}

\begin{prop}[$\mathrm{BP}(\mathbf{H}_E)$ better than $\mathrm{BP}(\mathbf{H}_B)$] \label{prop_supnb}
$\mathbb{S}_E \subset \mathbb{S}_B$.
\end{prop}
\begin{IEEEproof}
Suppose to the contrary that $\exists{\ }\mathcal{S} \in \mathbb{S}_E \setminus \mathbb{S}_B$.
Then, there is a parity-check equation in $\mathbf{H}_B$ that involves only one erasure, 
say $x_{i, e}^B = (\Phi_B(X_i))_e$. Let this equation be
\begin{align} 
\sum_{j \in \mathbb{I}_i}(\Phi_B(X_i))_j &+ \sum_{k \in \mathbb{I}} \sum_{j \in \mathbb{I}_k}(\Phi_B(X_k))_{j} = 0 \notag \\
\Rightarrow \sum_{j \in \mathbb{I}_i}x_{i, j}^B &+ \sum_{k \in \mathbb{I}} \sum_{j \in \mathbb{I}_k} x_{k, j}^B = 0 \label{eq_bbeq}
\end{align}
for some index sets $\mathbb{I}_i, \mathbb{I}_k \subset [b],{\ }\forall{\ }k \in \mathbb{I}, \mathbb{I} \subset [n_q] \setminus \{i\}$ and $e \in \mathbb{I}_i$. We know from 
the construction of the extended binary image that there exist variables 
\begin{align}
x_{k,\Sigma_k}^E &\triangleq (\Phi_E(X_k))_{\sum_{j \in \mathbb{I}_k}2^{j-1}}  = \sum_{j \in \mathbb{I}_{k}}(\Phi_B(X_k))_{j} \notag \\
&= \sum_{j \in \mathbb{I}_{k}}(\Phi_E(X_k))_{2^{j-1}}, k \in \mathbb{I} \cup \{i\}. \notag
\end{align}
From Lemma 
\ref{lem:EB-has-BB-eqns}, we also know that the equation 
\begin{equation} \label{eq_ebeq}
x_{i,\Sigma_i}^E + \sum_{k \in \mathbb{I}} x_{k,\Sigma_k}^E = 0
\end{equation}
is contained in $\mathbf{H}_{PM}$. Since every bit except $x_{i, e}^B$ was known in Equation \eqref{eq_bbeq}, 
every bit except $x_{i,\Sigma_i}^E $ is known in Equation \eqref{eq_ebeq}. Hence the extended binary image 
can solve for $x_{i,\Sigma_i}^E$. But the simplex portion of $\mathbf{H}_E$ contains the equation
\[
x_{i,\Sigma_i}^E = \sum_{j \in \mathbb{I}_i} x_{i, 2^{j - 1}}^E = \sum_{j \in \mathbb{I}_i}(\Phi_B(X_i))_j
\]
which has only one unknown $x_{i, 2^{e - 1}}^E = x_{i, e}^B$, which can also be solved for. This is a 
contradiction to the assumption that $\mathcal{S} \in \mathbb{S}_E$. Therefore, 
$\mathbb{S}_E \setminus \mathbb{S}_B = \emptyset$.
\end{IEEEproof}

\begin{eg} \label{eg_2}
Consider the code in Example \ref{eg_1} and let the received word be 
$(?0?{\ }0?0{\ }000)$, where $?$ denotes an erasure. It is easy to see that 
this received word corresponds to a stopping set of the basic binary image. 
However, the extended binary image can recover all erasures. Using $P_6$ we 
can recover $x_{2,7}^E$, then using the simplex equation 
$x_{2,7}^E = x_{2,1}^E + x_{2,2}^E + x_{2,4}^E$ we obtain $x_{2,2}^E$. In turn, 
$x_{2,i}^E \, \forall i \in \{1, \ldots, 7\}$ can be obtained. Finally we can use $P_3$ and $P_7$ to 
recover $x_{1,1}^E$ and $x_{1,4}^E$.
\end{eg}

Note that although we have only shown improper containment of $\mathbb{S}_E$ in 
the set of stopping sets of $\mathbf{H}_B$, in most cases this containment is proper, i.e., 
$\mathbb{S}_E \subsetneq \mathbb{S}_B$. One case where $\mathbb{S}_E = \mathbb{S}_B$ 
is when the Tanner graph corresponding to $\mathbf{H}_B$ is cycle-free.

Proposition \ref{prop_supnb} gives us an analytical insight into why non-binary codes perform well. However, since the 
superiority is established only in comparison with the basic binary 
image corresponding to the non-binary code, the result constitutes only a partial answer to why non-binary 
codes are better than their binary counterparts in general.

\section{Enhancing $\mathcal{C}_{B}$ using Redundant Parity-Checks} \label{sec_enh}
It is known that the performance of a linear code over the BEC with BP can be improved by using 
a parity-check matrix with redundant parity-checks (RPCs). In fact, by adding enough parity-checks, 
we can guarantee ML performance with BP (See \cite{han_08_isit_mlred} and references therein). A similar 
notion is that of \emph{stopping redundancy} \cite{sch_06_tit_stopred}, where RPCs are added to remove 
stopping sets of size smaller than the minimum distance of the code. In the same spirit, we pose the 
question whether it is possible to achieve the performance of a non-binary code by its basic binary 
image with some RPCs.

\begin{prop}
For every $\mathcal{S} \in \mathbb{S}_B \setminus \mathbb{S}_E$, there exists an RPC for the basic binary image,
$\vect{h}_\mathcal{S} \otimes_2 \Phi_B(\vect{X})^\mathsf{T} = \vect{0}$, such that 
$\mathcal{S} \notin \mathbb{S}(\hat{\mathbf{H}}_B)$ where 
$\hat{\mathbf{H}}_B^{\mathsf{T}} = (\mathbf{H}_B^{\mathsf{T}} | \vect{h}_\mathcal{S}^{\mathsf{T}})$.
\end{prop}
\begin{IEEEproof}
Since $\mathcal{S} \notin \mathbb{S}_E$, the BP decoder working over $\mathbf{H}_E$ can solve for some 
erased transmitted bit $x_{i,2^{e-1}}^E = x_{i,e}^B$. This implies that the ML decoder working with $\mathbf{H}_B$ 
can also solve for $x_{i,e}^B$. Since ML decoding over the BEC is the same as Gaussian elimination, the 
above means that there exists a linear combination of the parity-check equations in $\mathbf{H}_B$ 
which has only $x_{i,e}^B$ as the unknown, which can be set as $\vect{h}_\mathcal{S}$.
\end{IEEEproof}
Note that since there might be multiple linear combinations of parity-check equations of $\mathbf{H}_B$ 
that can solve for $x_{i,e}^B$, the RPC $\vect{h}_\mathcal{S}$ is not always unique.

We now describe an algorithm that finds the RPC $\vect{h}_\mathcal{S}$, pseudocode for which is given as Algorithm \ref{algo}.
\begin{algorithm}
\caption{Algorithm to find $\vect{h}_\mathcal{S}$ given $\mathcal{S} \in \mathbb{S}_B \setminus \mathbb{S}_E$
}
\label{algo}
\begin{algorithmic}[1]
\STATE $\mathcal{R} \gets \emptyset$, $\mathcal{L} \gets \emptyset$, $\vect{h}_{\mathcal{S}} \gets \vect{0}$
\WHILE {$\exists \text{ a row } \vect{h}_j^E \in \mathbf{H}_E$ that solves $(\Phi_E(\vect{X}))_i$}
  \STATE $\mathcal{R} \gets \mathcal{R} \cup \{(i,j)\}$
    \IF {$i \in \mathcal{I}_t$}
      \STATE $\mathcal{L} \gets \{(i,j)\}$ 
      \WHILE {$\mathcal{L} \neq \emptyset$}
	\STATE $(i,j) \gets \text{pop}(\mathcal{L})$
	\STATE Let $\mathbb{I}_j \subset [(2^b-1)n_q]: \sum_{k \in \mathbb{I}_j} \vect{u}_{k,(2^b-1)n_q} = \vect{h}_j^E$
	\FORALL {$i^\prime \in \mathbb{I}_j \cap \mathcal{I}_p \setminus \{i\}$}
	  \STATE Locate $(i^\prime, j^\prime) \in \mathcal{R}$
	  \STATE push$(\mathcal{L}, (i^\prime, j^\prime))$
	\ENDFOR
	\FORALL {$k \in \mathbb{I}_j \cap \mathcal{I}_t$}
	  \STATE $\sigma_k \gets \lfloor\frac{k}{2^b-1}\rfloor b + (k \mod (2^b-1))$
	  \STATE $\vect{h}_\mathcal{S} \gets \vect{h}_\mathcal{S} \oplus_2 \vect{u}_{\sigma_k, bn_q}$
	\ENDFOR
      \ENDWHILE
      \RETURN $\vect{h}_\mathcal{S}$
      \STATE {\bf terminate}
    \ENDIF
\ENDWHILE

  
\end{algorithmic}
\end{algorithm}
Given $\mathcal{S}$, the algorithm starts the peeling decoding over $\mathbf{H}_E$ with the erasure pattern 
$\mathcal{S} \cup \mathcal{I}_p$. The decoder 
maintains a list
of all the punctured bits that were solved 
and the unerased transmitted bits that were used to solve them. When a transmitted but erased
bit is solved, the decoder is terminated and the algorithm finds an equation involving only this recovered transmitted bit and some unerased transmitted bits chosen based on the punctured bits that appear in the computation tree for this recovered transmitted bit.
Thus, for a given stopping set, the algorithm uses a partial peeling decoding attempt over the extended binary image and a traversal through the computation tree of a recoverable transmitted bit 
to 
obtain the corresponding RPC of $\mathbf{H}_B$.
For a given collection of stopping sets, this algorithm is first run for low-weight stopping sets since RPCs that eliminate low-weight stopping sets may also eliminate higher weight stopping sets containing those low-weight stopping sets as a subset. It is possible to optimize the choice of RPCs to minimize the degrees of the additional check nodes introduced by them. However, this optimization was not considered in the implementation used for this paper.
 Note that the idea here is different from the one 
in \cite{sy_11_arx_extbinimg} where the authors try to optimize the set of transmitted bits under the assumption that 
bits of the extended binary image indexed by a subset of $\mathcal{I}_p$ are also transmitted to 
obtain a lower rate code that performs better.

Tables \ref{tab_ssw_4} and \ref{tab_ssw_8} list
the number of codewords and number of stopping sets for two non-binary 
LDPC codes, $\mathcal{C}_4$ and $\mathcal{C}_8$, and their binary images. Tanner graph for $\mathcal{C}_4$ is constructed randomly with parameters $q = 2^2, n_q = 96, k_q = 32$, and 
$d_l = 2, d_r = 3$. Tanner graph for $\mathcal{C}_8$ is constructed using Progressive Edge-Growth (PEG) \cite{Hu_03_tit_regular_peg} with parameters $q = 2^3, n_q = 100, k_q = 50 $, and $d_l = 2, d_r \simeq 4$. Let $A_B^w$ denote the number of codewords of weight $w$ in the basic binary image $\mathcal{C}_B$. Let $\mathbb{S}^w(\mathbf{H})$ denote the set of 
stopping sets of $\mathbf{H}$ of weight $w$, $\mathbb{S}_E^w$ the set of stopping sets of the extended 
binary image of weight $w$.
For $\mathcal{C}_4$, RPCs were added to its basic binary image to remove extra stopping sets of weights up to $12$ 
in $\mathbf{H}_B^{12}$ and weights up to $15$ in $\mathbf{H}_B^{15}$
while, for $\mathcal{C}_8$, RPCs were added to remove extra stopping sets of weights up to $8$ 
in $\mathbf{H}_B^{8}$ and weights up to $10$ in $\mathbf{H}_B^{10}$.
The stopping sets $\mathbb{S}_B = \mathbb{S}(\mathbf{H}_B)$ were found using the algorithm in \cite{ros_09_tit_stopset} and whether these were 
in $\mathbb{S}_E$ was verified by running the BP decoder over $\mathbf{H}_E$. For those stopping sets in $\mathbb{S}_B \setminus \mathbb{S}_E$, 
the RPCs were found using the algorithm described earlier in this section.
For $\mathcal{C}_4$, the basic binary image has 192 variable nodes and 128 check nodes, while 
the number of RPCs in $\mathbf{H}_B^{12}$ and $\mathbf{H}_B^{15}$ are $93$ and $154$, respectively. 
For comparison, the extended binary image has 288 variable nodes and 288 check nodes.
Similarly, the basic binary image for $\mathcal{C}_8$ has 300 variable nodes and 150 check nodes, while the number of RPCs in $\mathbf{H}_B^{8}$ 
and $\mathbf{H}_B^{10}$ is $60$ and $185$, respectively. The number of variable and check nodes in the extended binary image for $\mathcal{C}_8$ is $700$ and $1050$, respectively.
In general, for a NB-LDPC code over $\mathrm{GF}(q)$, the number of variable and check nodes in the basic binary image is $O(\log q)$ while it is $O(q)$ in the extended binary image. The number of variables nodes in the enhanced basic binary image is the same as basic binary image, while the number of check nodes is the sum of the number of check nodes in the basic binary image and the additional RPCs added.

\begin{table}[htbp]
\caption{Stopping set weights for a $\mathrm{GF}(4)$ code, $\mathcal{C}_4$, and its binary images}
\vspace{-2mm}
\centering
\begin{tabular}{cccccc}
\vspace{1mm}$w$ & $A_B^w$ & $|\mathbb{S}_E^w|$ & $|\mathbb{S}^w(\mathbf{H}_B)|$ & $|\mathbb{S}^w(\mathbf{H}_B^{12})|$ & $|\mathbb{S}^w(\mathbf{H}_B^{15})|$ \\
\hline 
\hline
$\leq$ $4$ & $0$ & $0$ & $0$ & $0$ & $0$ \\ 
$5$ & $4$ & $4$ & $6$ & $4$ & $4$ \\ 
$6$ & $2$ & $2$ & $11$ & $2$ & $2$ \\ 
$7$ & $7$ & $7$ & $31$ & $7$ & $7$ \\
$8$ & $7$ & $9$ & $50$ & $9$ & $9$ \\
$9$ & $3$ & $4$ & $86$ & $4$ & $4$ \\
$10$ & $12$ & $23$ & $171$ & $23$ & $23$ \\ 
$11$ & $22$ & $30$ & $343$ & $30$ & $30$ \\ 
$12$ & $63$ & $80$ & $873$ & $80$ & $80$ \\ 
$13$ & $87$ & $120$ & $2199$ & $151$ & $120$ \\ 
$14$ & $122$ & $204$ & $5463$ & $287$ & $204$ \\
$15$ & $205$ & $418$ & $13891$ & $650$ & $418$ \\ 
$16$ & $369$ & $793$ & $35209$ & $1322$ & $821$ \\
\hline
\end{tabular}
\label{tab_ssw_4}
\vspace{-1mm}
\end{table}

\begin{table}[htbp]
\caption{Stopping set weights for a $\mathrm{GF}(8)$ code, $\mathcal{C}_8$, and its binary images}
\vspace{-2mm}
\centering
\begin{tabular}{cccccc}
\vspace{1mm}$w$ & $A_B^w$ & $|\mathbb{S}_E^w|$ & $|\mathbb{S}^w(\mathbf{H}_B)|$ & $|\mathbb{S}^w(\mathbf{H}_B^{8})|$ & $|\mathbb{S}^w(\mathbf{H}_B^{10})|$ \\
\hline 
\hline
$\leq$ $2$ & $0$ & $0$ & $0$ & $0$ & $0$ \\ 
$3$ & $0$ & $0$ & $9$ & $0$ & $0$ \\ 
$4$ & $0$ & $0$ & $0$ & $0$ & $0$ \\
$5$ & $0$ & $0$ & $0$ & $0$ & $0$ \\ 
$6$ & $1$ & $1$ & $59$ & $1$ & $1$ \\ 
$7$ & $5$ & $5$ & $82$ & $5$ & $5$ \\
$8$ & $13$ & $13$ & $509$ & $13$ & $13$ \\
$9$ & $38$ & $38$ & $2781$ & $156$ & $38$ \\
$10$ & $64$ & $64$ & $11763$ & $612$ & $64$ \\ 
$11$ & $143$ & $147$ & $45310$ & $2102$ & $310$ \\ 
$12$ & $309$ & $358$ & $169120$ & $6363$ & $871$ \\ 
$13$ & $799$ & $970$ & $663617$ & $19346$ & $2575$ \\ 
$14$ & $1906$ & $2525$ & $2727519$ & $62032$ & $7893$ \\
\hline
\end{tabular}
\label{tab_ssw_8}
\vspace{-1mm}
\end{table}

\begin{figure}[!h]
\centering
\includegraphics[scale=0.985]{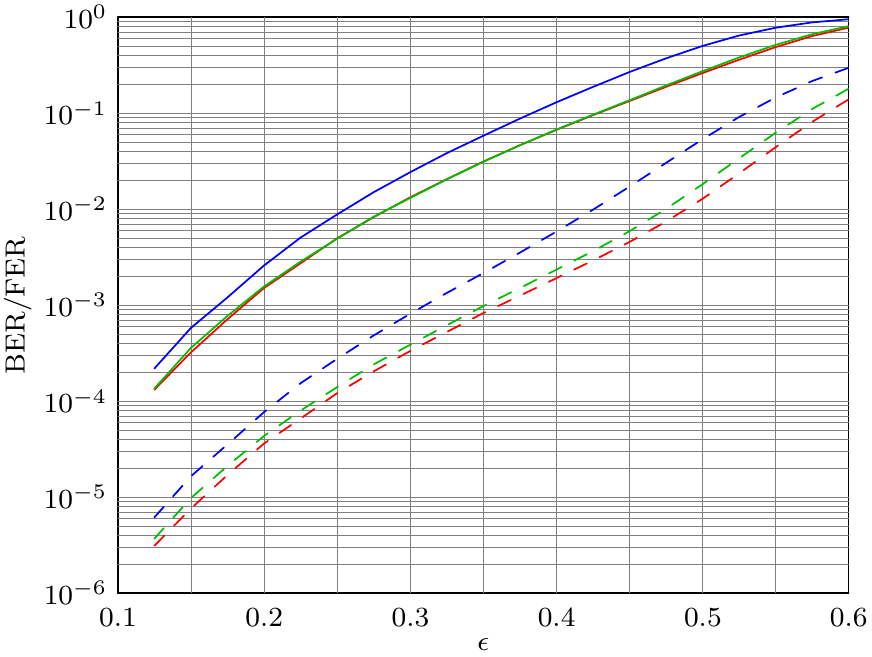} 
\begin{picture}(0.1,0.1)(45,-24){\includegraphics[scale=0.32]{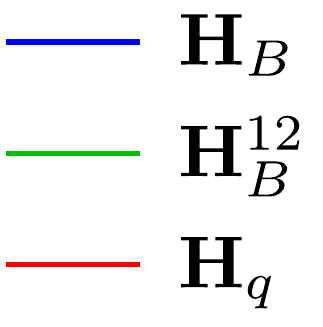}}\end{picture}   
\vspace{-4mm}
\caption{Bit (dashed curves) and frame (solid curves) error rates for the random 
non-binary code $\mathcal{C}_4$ with parameters $q = 2^2, n_q = 96, k_q = 32, d_l = 2, d_r = 3$ and its binary images considered in Table \ref{tab_ssw_4}. 
\vspace{-2mm}}
\label{fig_sim_4}
\end{figure}

\begin{figure}[!h]
\centering
\includegraphics[scale=0.985]{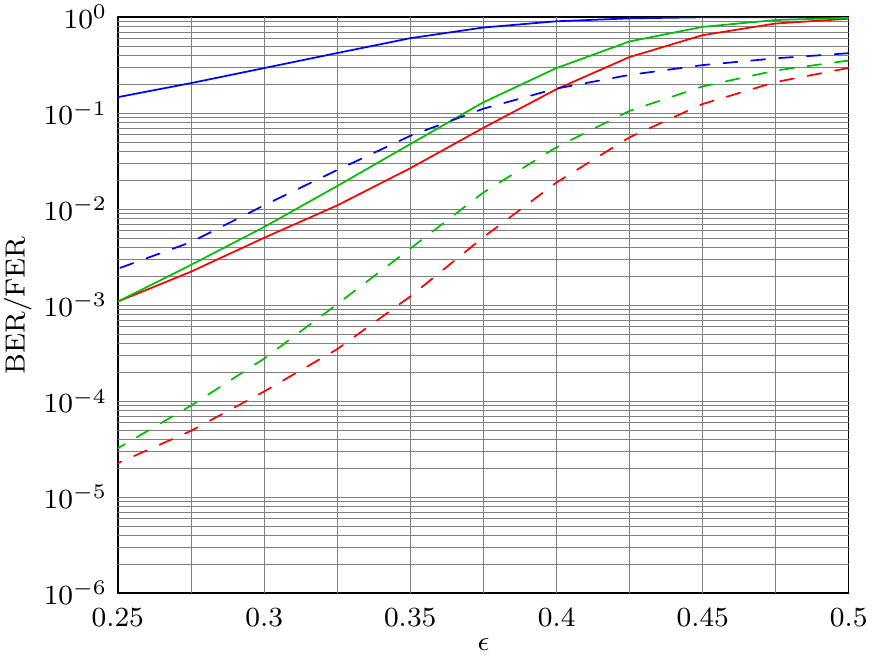}  
\begin{picture}(0.1,0.1)(45,-24){\includegraphics[scale=0.32]{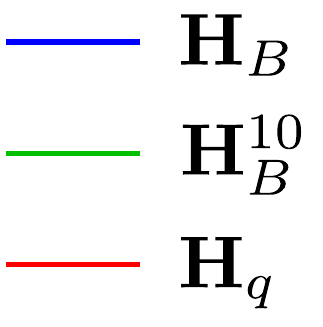}}\end{picture} 
\vspace{-4mm}
\caption{Bit (dashed curves) and frame (solid curves) error rates for the PEG constructed
non-binary code $\mathcal{C}_8$ with parameters $q = 2^3, n_q = 100, k_q = 50, d_l = 2, d_r \simeq 4$ and its binary images considered in Table \ref{tab_ssw_8}. 
\vspace{-2mm}}
\label{fig_sim_8}
\end{figure}

For the codes under consideration, Figures \ref{fig_sim_4} and \ref{fig_sim_8} plot the BP performance of the non-binary codes, their basic binary images, and the enhanced binary images.
The improvement in performance with RPCs is evident in the figures. The FERs of the non-binary 
code and the enhanced binary image are close to each other, and the corresponding BERs are also 
comparable.

\section{Conclusion} \label{sec_conc}
We showed that when the transmission occurs over the BEC, BP decoding over the non-binary graph 
has a better performance than BP decoding over the Tanner graph of the basic binary image of the 
code. 
We proposed an algorithm to efficiently find effective redundant parity-checks for the basic binary
image by observation of the BP decoding iterations of the extended binary image.
Through numerical results and simulations, the effectiveness of the proposed algorithm was established. 
Obtaining bounds on the number of RPCs of the basic binary image needed to achieve the same performance as BP over non-binary code would be of interest.
Similar characterization of the reasons for the superiority of
non-binary codes over other channels involving errors as well as erasures
will give further insight on designing strong codes and efficient decoding
algorithms for such channels.

\section{Acknowledgment}
This work was funded in part by NSF Grant CCF-0829865 and by Western Digital. The authors would like to thank Xiaojie Zhang for 
help with simulations.

\twobibs{
\bibliographystyle{IEEEtran}
}
{

}

\end{document}